\documentclass{cjpsuppl}% for LaTeX 2e
\usepackage{graphicx}
\usepackage{cite}
\usepackage{epsfig}

\begin{document}
\title{Probing extra dimensions through the
invisible Higgs decay}
\authori{D.\,Dominici}
\addressi{Dipartimento di Fisica, Universit\`a di Firenze,  I-50019 Sesto
F., Italy\\
I.N.F.N.,
Sezione di Firenze,  I-50019 Sesto F., Italy\\
}
\authorii{}    \addressii{}
\authoriii{}   \addressiii{}
\authoriv{}    \addressiv{}
\authorv{}     \addressv{}
\authorvi{}    \addressvi{}
\headtitle{Probing extra dimensions through the
invisible Higgs decay \ldots}
\headauthor{D. Dominici}
\lastevenhead{D. Dominici:  \ldots}
%%%%%%%%%%%%%% Pro editory supplementu: %%%%%%%%%%%%%%%
\refnum{}%slouzi editorum pro evidenci; nakonec {}
\daterec{xxx;\\final version xxx}
\suppl{A}  \year{2003} \setcounter{page}{1}
%\firstpage{1}
%\lastpage{000}
%\makefirsttitle
%%%%%%%%%%%%%%%%%%%%%%%%%%%%%%%%%%%%%%%%%%%%%%
\maketitle

\def\lsim{\mathrel{\raise.3ex\hbox{$<$\kern-.75em\lower1ex\hbox{$\sim$}}}}
\def\gsim{\mathrel{\raise.3ex\hbox{$>$\kern-.75em\lower1ex\hbox{$\sim$}}}}
\def\md{M_D}
\def\bit{\begin{itemize}}
\def\eit{\end{itemize}}
\def\nsd{N_{SD}}
\def\rts{\sqrt s}
\def\ie{{\it i.e.}}
\def\fbi{~{\rm fb}^{-1}}
\def\fb{~{\rm fb}}
\def\mw{m_W}
\def\slasht#1{#1\hskip-8pt/\hskip5pt}
\def\mpl{\overline M_{Pl}}
\def\to{\rightarrow}
\def\ptl{\partial}
\def\beq{\begin{equation}}
\def\eeq{\end{equation}}
\def\bea{\begin{eqnarray}}
\def\eea{\end{eqnarray}}
\def\nn{\nonumber}
\def\half{{1\over 2}}
\def\rhalf{{1\over \sqrt 2}}
\def\calo{{\cal O}}
\def\cala{{\cal A}}
\def\call{{\cal L}}
\def\calm{{\cal M}}
\def\del{\delta}
\def\eps{\epsilon}
\def\lam{\lambda}
\def\anti{\overline}
\def\delfac{\sqrt{{2(\del-1)\over 3(\del+2)}}}
\def\heff{h_{eff}}
\def\hp{h'}
\def\mheff{m_{h_{eff}}}
\def\square{\boxxit{0.4pt}{\fillboxx{7pt}{7pt}}\hspace*{1pt}}
    \def\boxxit#1#2{\vbox{\hrule height #1 \hbox {\vrule width #1
             \vbox{#2}\vrule width #1 }\hrule height #1 } }
    \def\fillboxx#1#2{\hbox to #1{\vbox to #2{\vfil}\hfil}   }

\def\gev{~{\rm GeV}}
\def\mev{~{\rm MeV}}
\def\tev{~{\rm TeV}}
\def\gam{\gamma}
\def\sn{s_{\vec n}}
\def\sm{s_{\vec m}}
\def\svk{s_{\vec k}}
\def\svl{s_{\vec l}}
\def\mm{m_{\vec m}}
\def\mn{m_{\vec n}}
\def\mk{m_{\vec k}}
\def\ml{m_{\vec l}}
\def\mh{m_h}
\def\mhp{m_{h'}}
\def\sumn{\sum_{\vec n>0}}
\def\summ{\sum_{\vec m>0}}
\def\vl{\vec l}
\def\vk{\vec k}
\def\ml{m_{\vl}}
\def\mk{m_{\vk}}
\def\etmiss{\slasht E_T}
\def\ptmiss{\slasht p_T}

\begin{abstract}
In the large extra dimension model of Arkani-Hamed, Dimopoulos and Dvali
the presence of an interaction between the Ricci scalar curvature and the
Higgs doublet of the Standard Model makes a
light Higgs boson observable at LHC at the $5~\sigma$ level  through 
the fusion process $pp\to W^*W^* +X \to Higgs,graviscalars
+X \to invisible+X$ 
for  the portion of the Higgs-graviscalar mixing ($\xi$) and
effective Planck mass ($M_D$) parameter space where channels relying
on visible Higgs decays fail to achieve a $5~\sigma$ signal.
However even if  the LHC has a good chance of seeing a signal, it will
not be able to determine the  parameters of the model
with any real precision. This goal can be reached by adding the
following   LC measurements:
$\gam+\etmiss$, Higgs production and decay in the 
visible SM-like final states and 
in the invisible final state. 
\end{abstract}

\section{Introduction}
The effect of the invisible decay of the Higgs on the Higgs phenomenology
at LHC has been recently considered. 
In several modifications of the Standard Model such a decay appears:
as invisible decay to neutralinos in supersymmetric models (for a recent analysis
see \cite{Belanger:2001am,Martin:1999qf}), as  decay to Majorons\cite{Joshipura:1993hp,Martin:1999qf} in models 
with spontaneously broken lepton number
or as a decay to neutrinos  in 
fourth generation models \cite{Belotsky:2002ym}.
The recent suggestion of a low scale quantum gravity (ADD)
\cite{Arkani-Hamed:1998rs,Antoniadis:1998ig}
has added a new mechanism for
predicting invisible Higgs decay, as  decay to Kaluza Klein neutrino
excitations \cite {Martin:1999qf} or to graviscalars \cite{Giudice:2000av,Wells:2002gq,Allanach:2004ub,lavoro}.
In ADD models the presence of an
interaction between the Higgs $H$ and the 
Ricci scalar curvature of the induced 4-dimensional metric $g_{ind}$, 
given by the following action
\be
S=-\xi \int d^4 x \sqrt{g_{ind}}R(g_{ind})H^\dagger H\, ,
\ee
generates, after the usual shift $H=({v+ h\over \sqrt{2}},0)$,
the following mixing term \cite{Giudice:2000av}
\begin{equation}
{\cal L}_{\rm mix}=\epsilon  h \sum_{\vec n >0}s_{\vec n}
\end{equation}
with
\beq
\eps=-{2\sqrt 2\over M_P}\xi v \mh^2\sqrt{{3(\del-1)\over \del+2}}\,.
\eeq
Above, $M_P=(8\pi G_N)^{-1/2}$ is the Planck mass, $\delta$ is the number of extra
dimensions, $\xi$ is a dimensionless parameter and
$s_{\vec n}$ is a graviscalar KK excitation with
mass $m_{\vec n}=2\pi\vert \vec n\vert /L$, $L$ being the
size of each of the extra dimensions.
 After diagonalization of the full mass-squared matrix
one finds that the physical eigenstate, $h'$,  acquires
admixtures of the graviscalar states and vice versa.
Dropping $\calo(\eps^2)$ terms and higher \cite{lavoro},
\beq
h'\sim \left[h-\sum_{\vec m>0}{\eps\over \mh^2-i \mh
\Gamma_{h}-m_{\vec m}^2}s_{\vec m}\right]\,,\quad
s'_{\vec m}\sim \left[ s_{\vec m}+{\eps\over \mh^2-i\mh\Gamma_{h} 
-m_{\vec m}^2} h\right]\,,
\label{eigenstate}
\eeq
where $\Gamma_h$ is the visible width.
In computing a process such as $WW\to h'+\sum_{\vec m>0}s_{\vec m}' \to F$,
normalization and mixing  corrections of order $\eps^2$ that 
are present must be taken into account and the full coherent sum
over physical states must be performed. The result at the
amplitude level is \cite{lavoro}
\beq
\cala(WW\to F)(p^2)\sim {g_{WWh}g_{h F}\over  
p^2-\mh^2+i\mh\Gamma_h+iG(p^2)+F(p^2)}\, ,
\label{amplitude}
\eeq
where 
$F(p^2)\equiv -\eps^2 {\rm Re} \left[\sum_{\vec m>0}{1\over
    p^2-m_{\vec m}^2}\right]$ and $G(p^2)\equiv
-\eps^2{\rm Im}\left[\sum_{\vec m>0}{1\over p^2-m_{\vec m}^2}\right]$.
Taking the amplitude squared and
integrating over $dp^2$ in the narrow width approximation
gives the result
\bea
\sigma(WW\to h'+\sum_{\vec m>0}s'_{\vec m}\to F)&=&\sigma_{SM}(WW\to h \to F)
\left[{1\over 1+F'(m_{h\,ren}^2)}\right]^2\nn\\
&&\times
\left[{\Gamma_h\over \Gamma_h+\Gamma_{h\to gravisc.}}\right]\, ,
\label{xsec}
\eea
where $m_{h\,ren}^2-\mh^2+F(m_{h\,ren}^2)=0$ and we have defined
$\mh\Gamma_{h\to gravisc.}\equiv G(m_{h\,ren}^2)$. 
For a light Higgs boson both the wave function renormalization
and the mass renormalization effects are small \cite{lavoro}. 
Therefore 
the coherently summed
amplitude gives the following result for the cross section:
\bea
\sigma(WW\to h'+\sum_{\vec m>0}s'_{\vec m}\to F)&\sim&\sigma_{SM}(WW\to h \to F)
\nn\\
&&\times
\left[{\Gamma_h\over \Gamma_h+\Gamma_{h\to gravisc.}}\right]\, ,
\label{xsecappr}
\eea
where  the invisible witdh is given by \cite{Giudice:2000av,Wells:2002gq,lavoro}
\begin{eqnarray}
\Gamma_{h\to gravisc.}&\equiv& \Gamma(h\to 
\sum_{\vec n>0}s_{\vec n})=
2\pi\xi^2 v^2 \frac {3(\delta -1)}
{\delta +2}
\frac {m_h^{1+\delta}}{M_D^{2+\delta}}{S_{\delta -1}}\nn\\
&\sim& (16\,MeV) 20^{\delta -2} \xi^2
S_{\delta-1}\frac {3(\delta -1)}
{\delta +2} \left ( \frac {m_h}{150\, GeV} \right )^{1+\delta}
\left ( \frac 
{3\, TeV} {M_D}\right )^{2+\delta}\, ,\nn\\
&&
\label{invwidth}
\end{eqnarray}
where $S_{\delta-1}=2\pi^{\delta/2}/\Gamma(\delta/2)$ denotes the
surface of a unit radius sphere in $\delta$ dimensions while $M_D$ is
related to the $D$ dimensional reduced Planck constant ${\overline
  M}_D$ by $M_D= (2\pi)^{\delta/(2+\delta)}{\overline M}_D$.

\begin{table}[htb]
\caption{95\% CL limits from Tevatron/LEP \cite{Giudice:2003tu}} \vspace{1mm}
\small
\begin{center}
\begin{tabular}{|c|c|c|c|c|c|}
%\small
\hline \raisebox{0mm}[4mm][2mm] {$\delta$} & 2 &
3 &4&5&6\\
%\multicolumn{2}{|c|} \\
\hline
%\multicolumn{1}{|l|}
\raisebox{0mm}[4mm] {$M_D$}  (TeV)& 1.45 & 1.09&0.87&0.72&0.65
\\
%L & M\\[0.5mm]
\hline
%\multicolumn{1}
\end{tabular}
\vspace{-1mm}
\end{center}
\label{leplimits}
\end{table}

 \begin{figure}[htb]
     \begin{center}
     \begin{tabular}{cc}
     \mbox{\epsfig{file=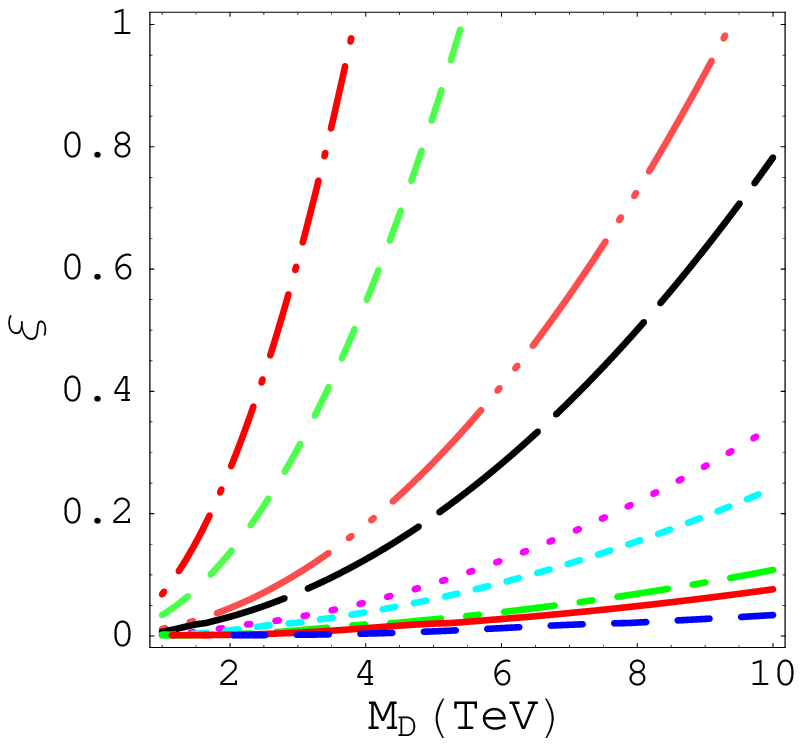,width=5.7cm}}&
     \mbox{\epsfig{file=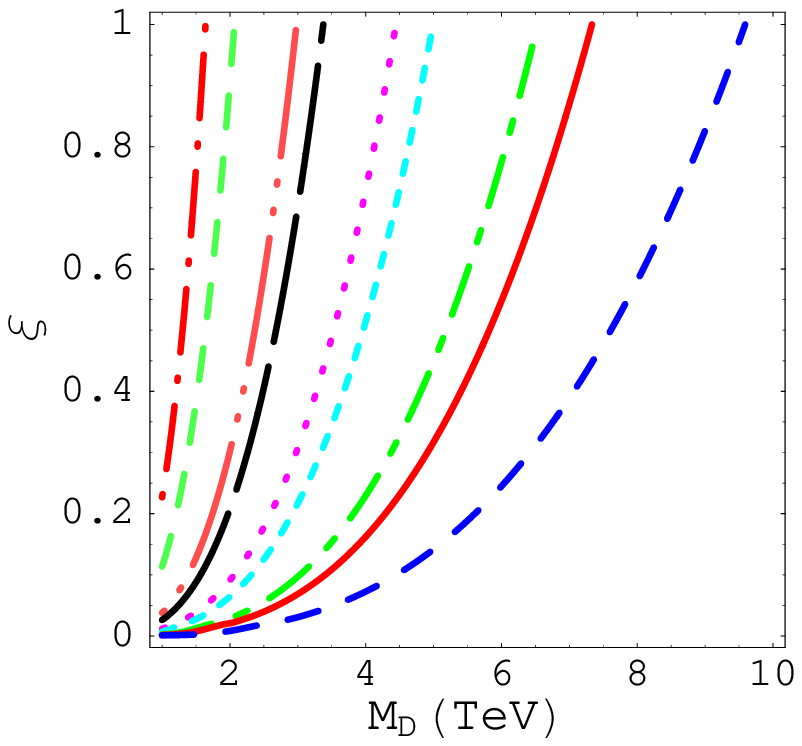 ,width=5.7cm}}\\
\mbox{\epsfig{file=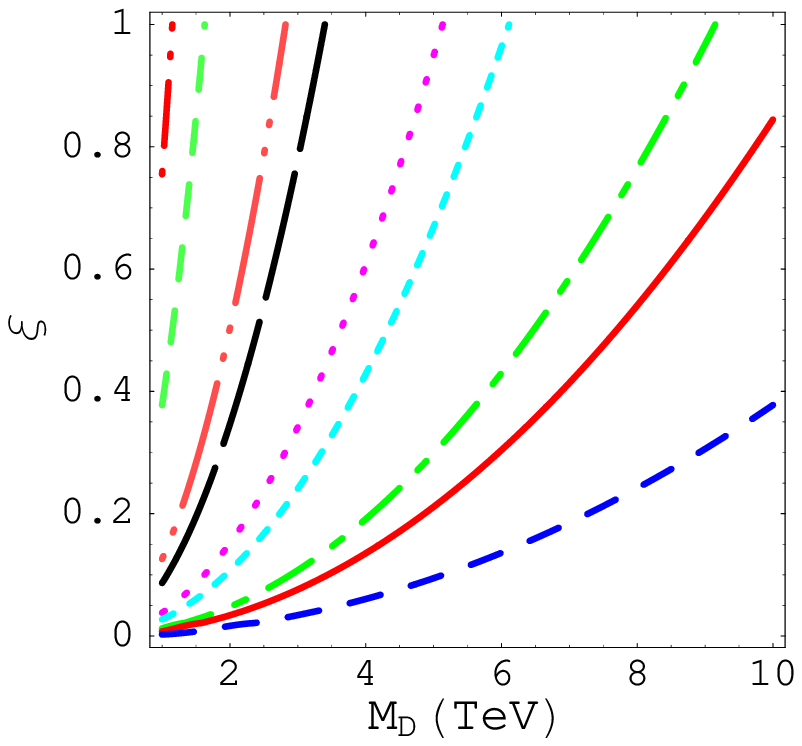,width=5.7cm}}&
     \mbox{\epsfig{file=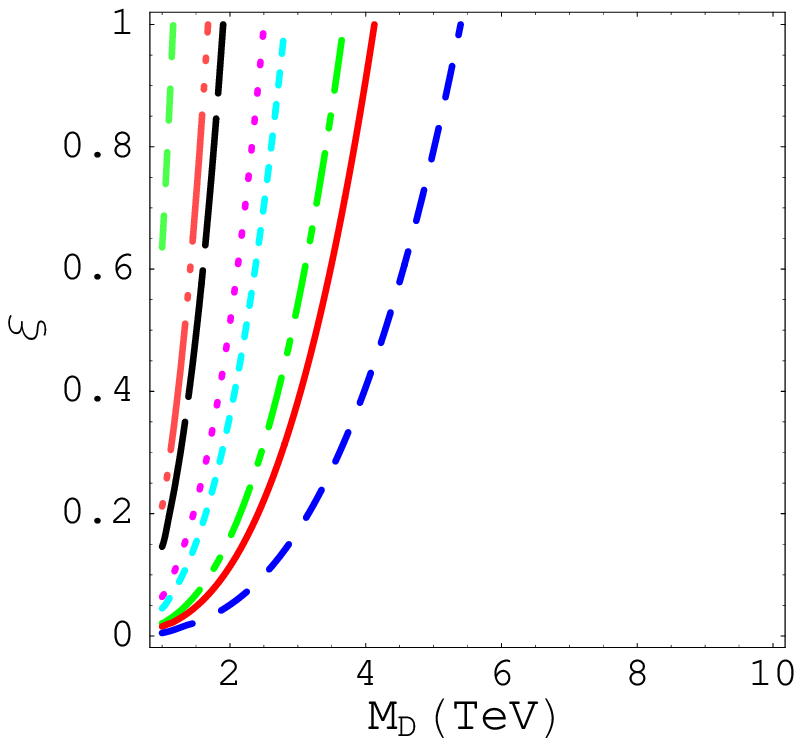 ,width=5.7cm}}
     \end{tabular}
     \end{center}
     \caption{Contours of fixed $BR(h\to inv)$  in the
$M_D$(TeV) -- $\xi$ parameter space for  $\del=2$ (left)
and $\del=4$ (right) in the upper part for   $m_h=120\gev$ and
in the lower part for $m_h=237\gev$. 
 In order of increasing $\xi$ values, the 
contours correspond to:
 $0.0001$
(large blue dashes), $0.0005$ (solid red line), $0.001$  (green
long dash -- short dash line), $0.005$ (short cyan dashes), 
$0.01$ (purple dots), $0.05$ (long black dashes), $0.1$ (chartreuse
long dashes with
double dots), $0.5$ (green dashes), and $0.85$ (red long dash,
short dot line at high $\xi$ and
low $M_D
$)
}
     \label{fig:br}
     \end{figure}

% \begin{figure}[htb]
%     \begin{center}
%     \begin{tabular}{cc}
%     \mbox{\epsfig{file=md_xi_widthcontours_del2_mh237.eps,width=5.7cm}}&
%     \mbox{\epsfig{file=md_xi_widthcontours_del4_mh237.eps ,width=5.7cm}}
%     \end{tabular}
%     \end{center}
%     \caption{Contours of fixed $BR(\heff\to graviscalar)$  in the
%$M_D$(TeV) -- $\xi$ parameter space for $m_h=237\gev$ $\del=2$ (left)
%and $\del=4$ (right).
% In order of increasing $\xi$ values, the width
%contours correspond to:
% $0.0001$
%(large blue dashes), $0.0005$ (solid red line), $0.001$ (green
%long dash -- short dash line), $0.005$ (short cyan dashes), 
%$.01$ (purple dots), $.05$ (long black dashes), $0.1$ (chartreuse
%long dashes with
%double dots),  $0.5$ (green dashes), and $0.85$ (red long dash,
%short dot line at high $\xi$ and
%low $M_D
%$)
%}
%     \label{fig:}
%     \end{figure}

\section{Detecting the Higgs at the LHC and LC}

\begin{figure}[htb]
\begin{center}
\begin{tabular}{c c}
\includegraphics[width=6.0cm,height=6.0cm]{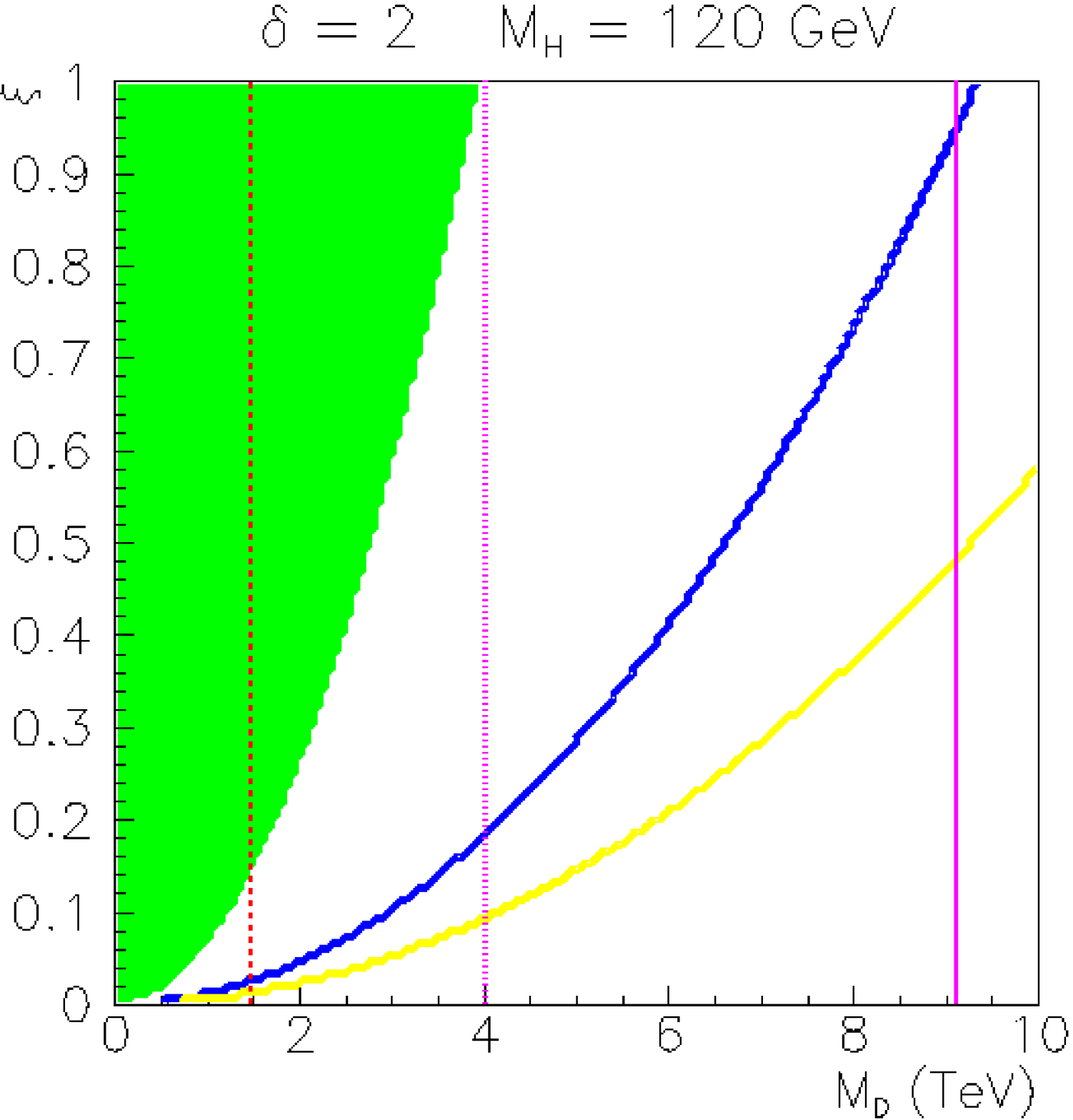} &
\includegraphics[width=6.0cm,height=6.0cm]{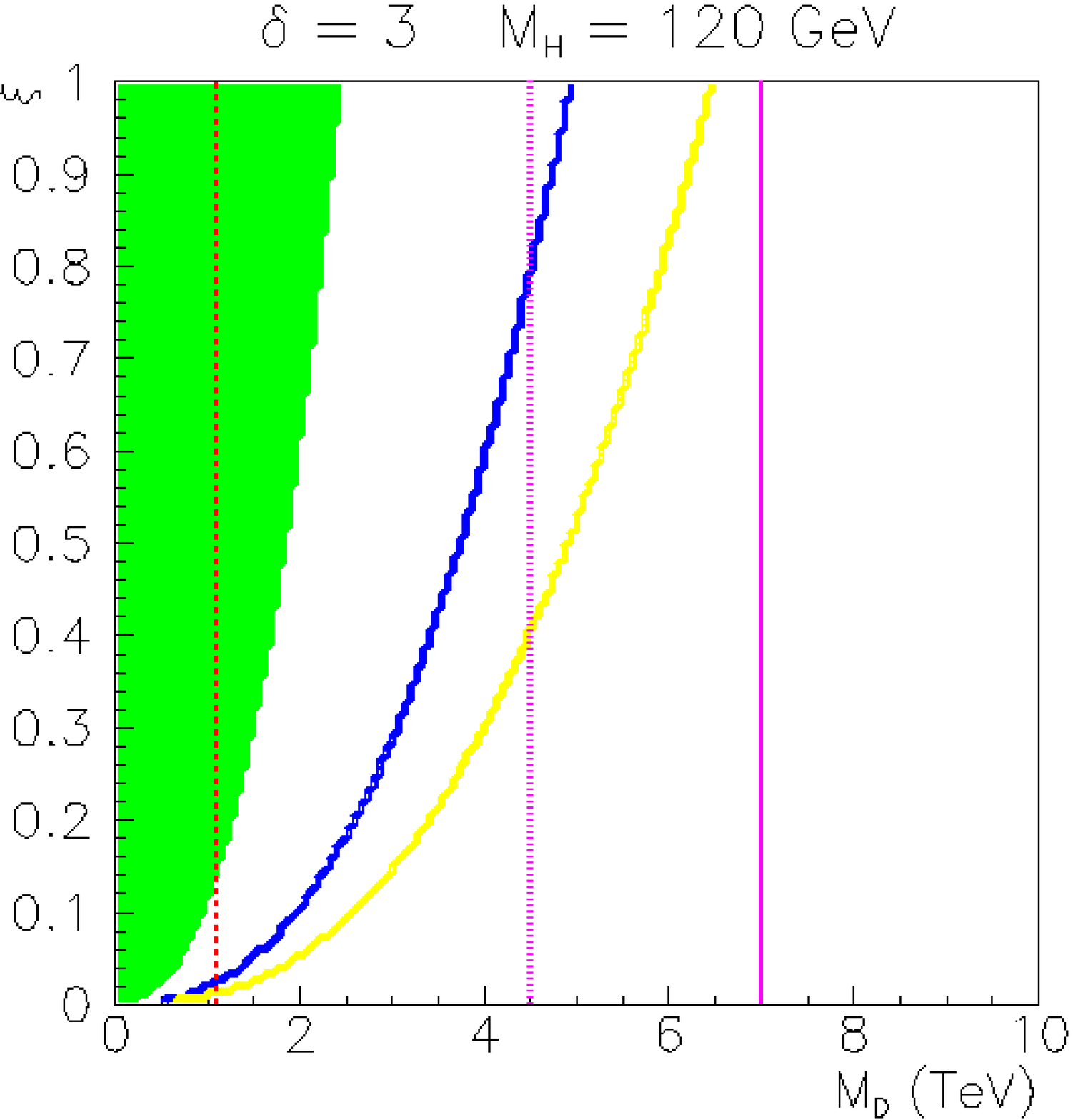}
\end{tabular}
\end{center}
\begin{center}
\begin{tabular}{c c}
\includegraphics[width=6.0cm,height=6.0cm]{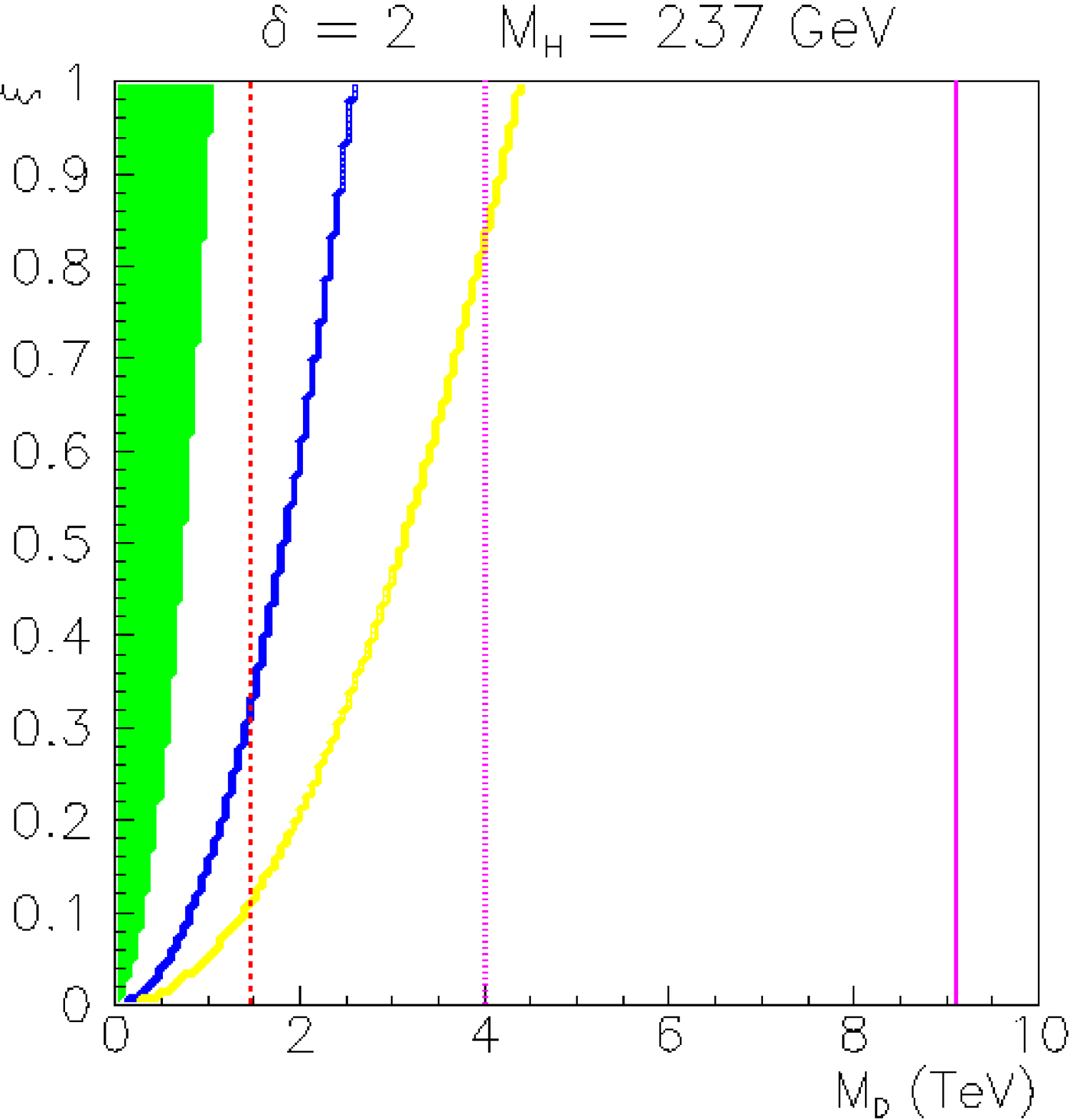}&
\includegraphics[width=6.0cm,height=6.0cm]{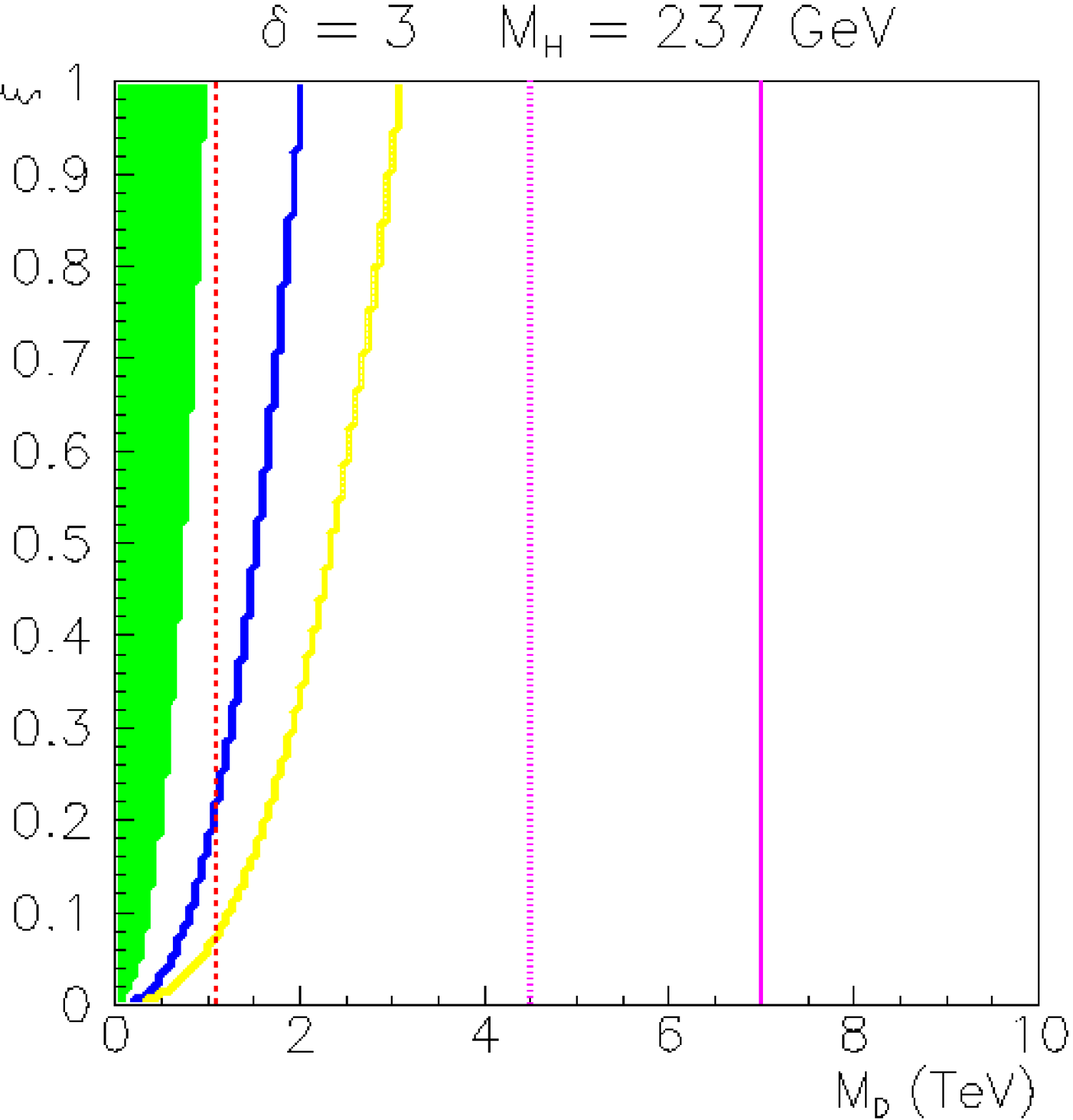}
\end{tabular}
\end{center}
\begin{center}
\caption{Invisible decay width effects in the $\xi$ - $M_D$ plane for 
  $m_h$ = 120~GeV (upper)
and  $m_h$ = 237~GeV (lower). The plots are for  $\delta$= 2
  (left) and $\delta$= 3 (right). The
  green (grey) regions indicate where the Higgs signal at the LHC
  drops below the 5~$\sigma$ threshold for 100 fb$^{-1}$ of data.  The
  regions above the blue (bold) line are the parts of the parameter
  space where the LHC invisible Higgs signal in the $WW$-fusion
  channel exceeds 5~$\sigma$ significance. The solid vertical line at
  the largest $M_D$ value in each figure shows the upper limit on
  $M_D$ which can be probed at the $5~\sigma$ level
  by the analysis of jets/$\gamma$ with
  missing energy at the LHC.  The middle dotted vertical line 
  shows the value of $M_D$ below which the
  theoretical computation at the LHC is ambiguous --- a signal could
  still be present there, but its magnitude is uncertain.   The dashed vertical line at the lowest $M_D$ value is
  the 95\% CL lower limit coming from combining Tevatron and LEP/LEP2 limits.
  The regions above the yellow (light grey) line are the parts of the
  parameter space where the LC invisible Higgs signal will exceed
  $5~\sigma$ assuming $\sqrt s=350\gev$ and $L=500\fbi$.  }
\label{fig:120}
\end{center}
%}
\end{figure}

Fig.~\ref{fig:br}  shows
that the branching ratio of the Higgs into invisible states
can be substantial  for $M_D$ values in the  TeV range both when
   $m_h=120$ GeV (upper part), therefore below the $WW$ threshold,  and 
 when $m_h=237$ GeV (lower part), a value  greater than the $WW$ threshold 
and corresponding to the 95\% CL limit from LEP data with $m_t=178$ GeV.
As a consequence 
this invisible
width causes a significant suppression of the LHC Higgs rate in the
standard visible channels and
for any given value of the Higgs boson mass, there is a considerable
parameter space region where the invisible decay width of the Higgs boson
could be the Higgs discovery channel.  This is exemplified in Figure~\ref{fig:120}
for $m_h$ = 120~GeV and $m_h$ = 237~GeV,  $\delta$= 2,3.
In the green (light grey) region the Higgs signal in
standard channels drops below the 5~$\sigma$ threshold with 100
fb$^{-1}$ of LHC data. But in the area above the bold blue line the
LHC search for invisible decays in the fusion channel yields a signal
with an estimated significance exceeding 5~$\sigma$. We have here rescaled
to higher luminosity 
the statistical significance of the analysis presented in \cite{cmsnote}. 
 The solid vertical line at
  the largest $M_D$ value in each figure shows the upper limit on
  $M_D$ which can be probed at the $5~\sigma$ level
  by the analysis of jets/$\gamma$ with
  missing energy at the LHC \cite{Vacavant:2001sd}.  The middle dotted vertical line 
  shows the value of $M_D$ below which the
  theoretical computation at the LHC is ambiguous --- a signal could
  still be present there, but its magnitude is uncertain \cite{Giudice:1998ck}.  
%The uncertainty comes from the fact that in this region the subprocess
%energy is higher than $M_D$.
The dashed vertical line at the lowest $M_D$ value is
  the 95\% CL lower limit coming from combining Tevatron and LEP/LEP2 limits
 (from Table \ref{leplimits}).
  The regions above the yellow (light grey) line are the parts of the
  parameter space where the LC invisible Higgs signal will exceed
  $5~\sigma$. 
We have employed the $\rts=350\gev$, $L=500\fbi$ results of
\cite{schumacher} looking for a peak in the $M_X$ mass spectrum
in $e^+e^-\to ZX$ events.

 In conclusion, 
whenever the Higgs boson sensitivity is lost due to
the suppression of the canonical decay modes, the invisible rate is
large enough to still ensure detection through the $WW$ fusion channel.

\begin{figure}[p]
\begin{center}
\begin{tabular}{c c}
%\epsfigwidth=6.0cm} &
%\epsfig{file=lhc_cont_3.eps,width=6.0cm}\\
%\epsfig{file=lhc_cont_4.eps,width=6.0cm} &
%\epsfig{file=lhc_cont_5.eps,width=6.0cm} \\
\includegraphics[width=4.5cm,height=6.5cm]{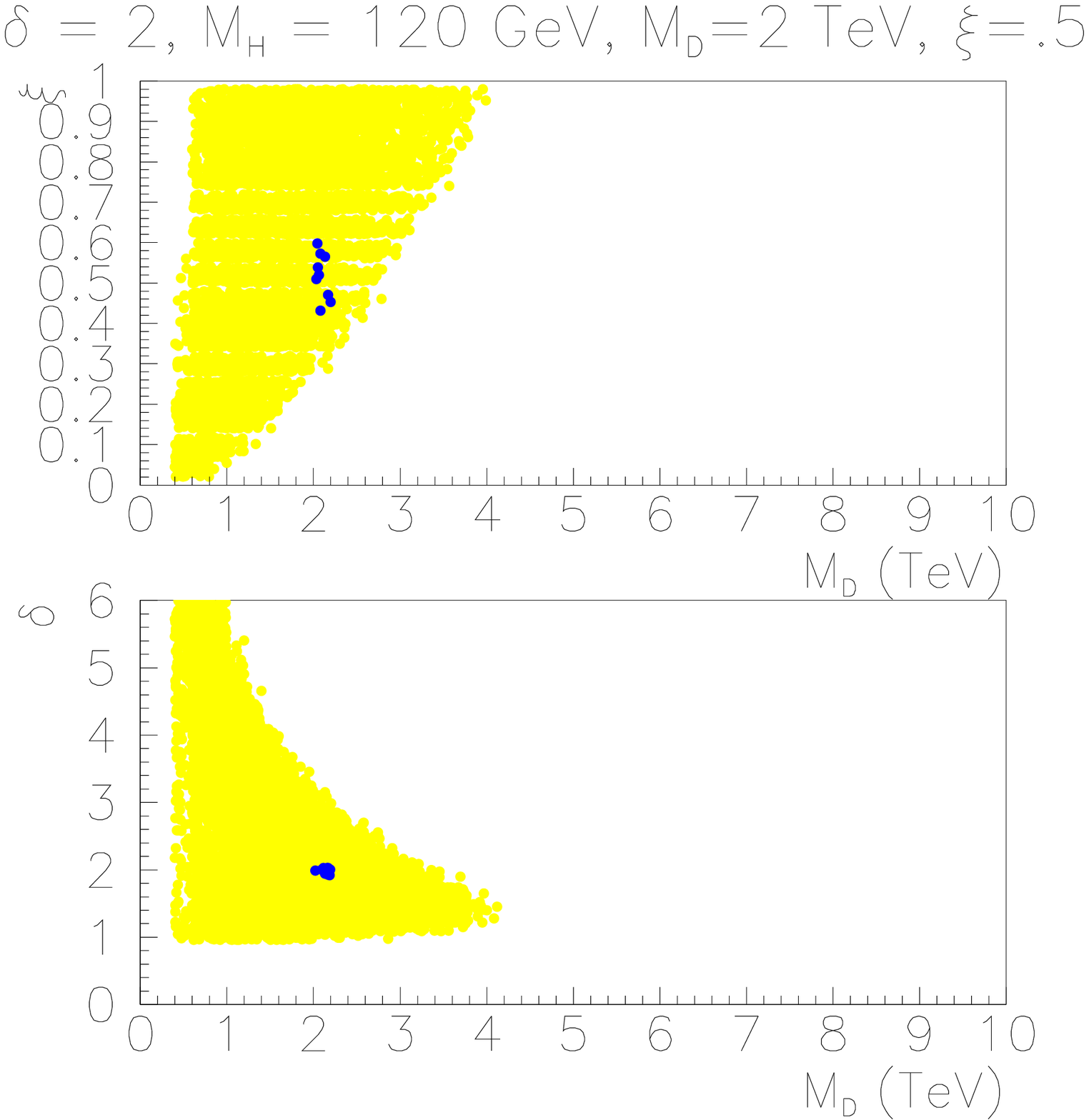} &
\includegraphics[width=4.5cm,height=6.5cm]{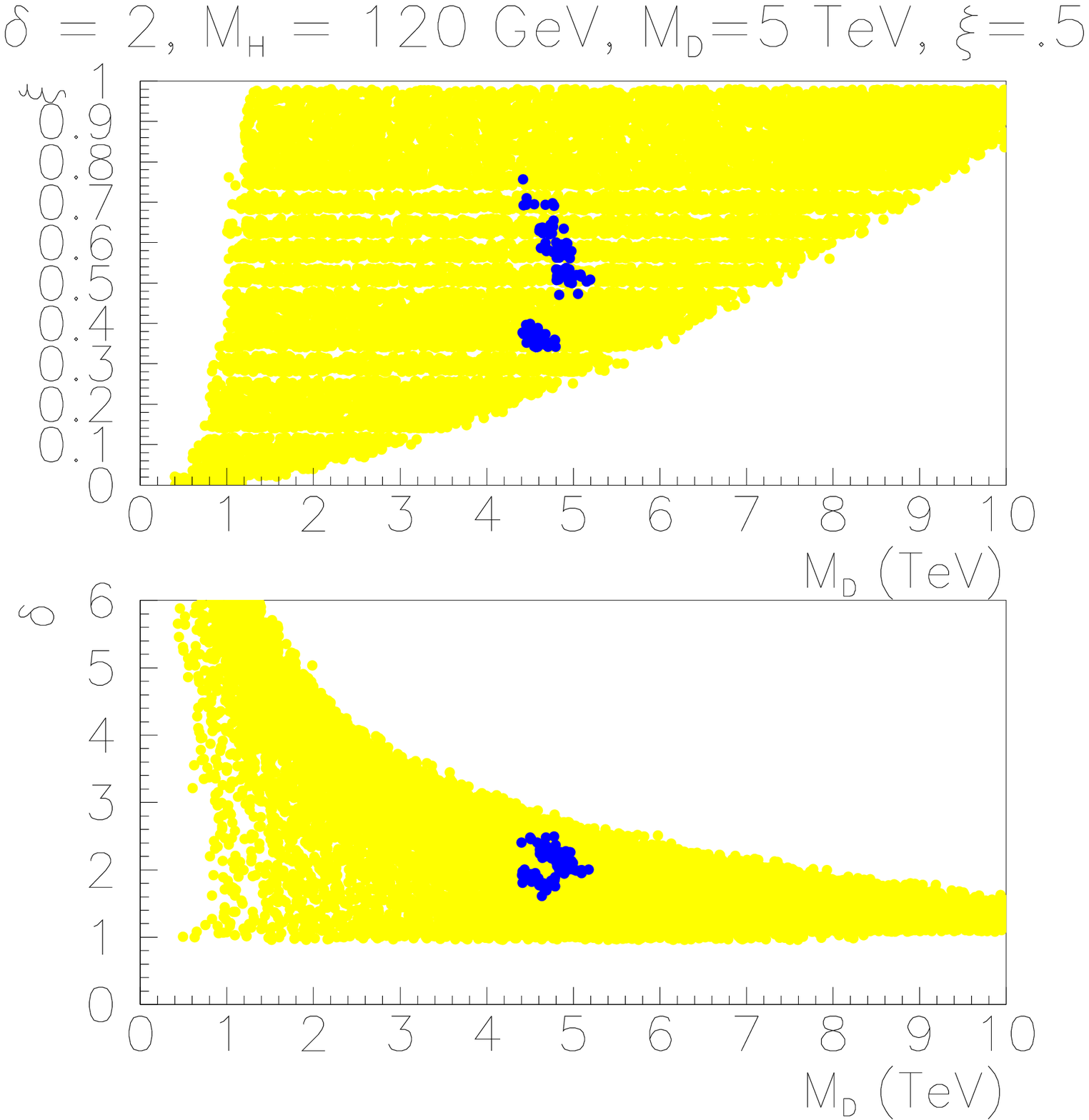}\\
\includegraphics[width=4.5cm,height=6.5cm]{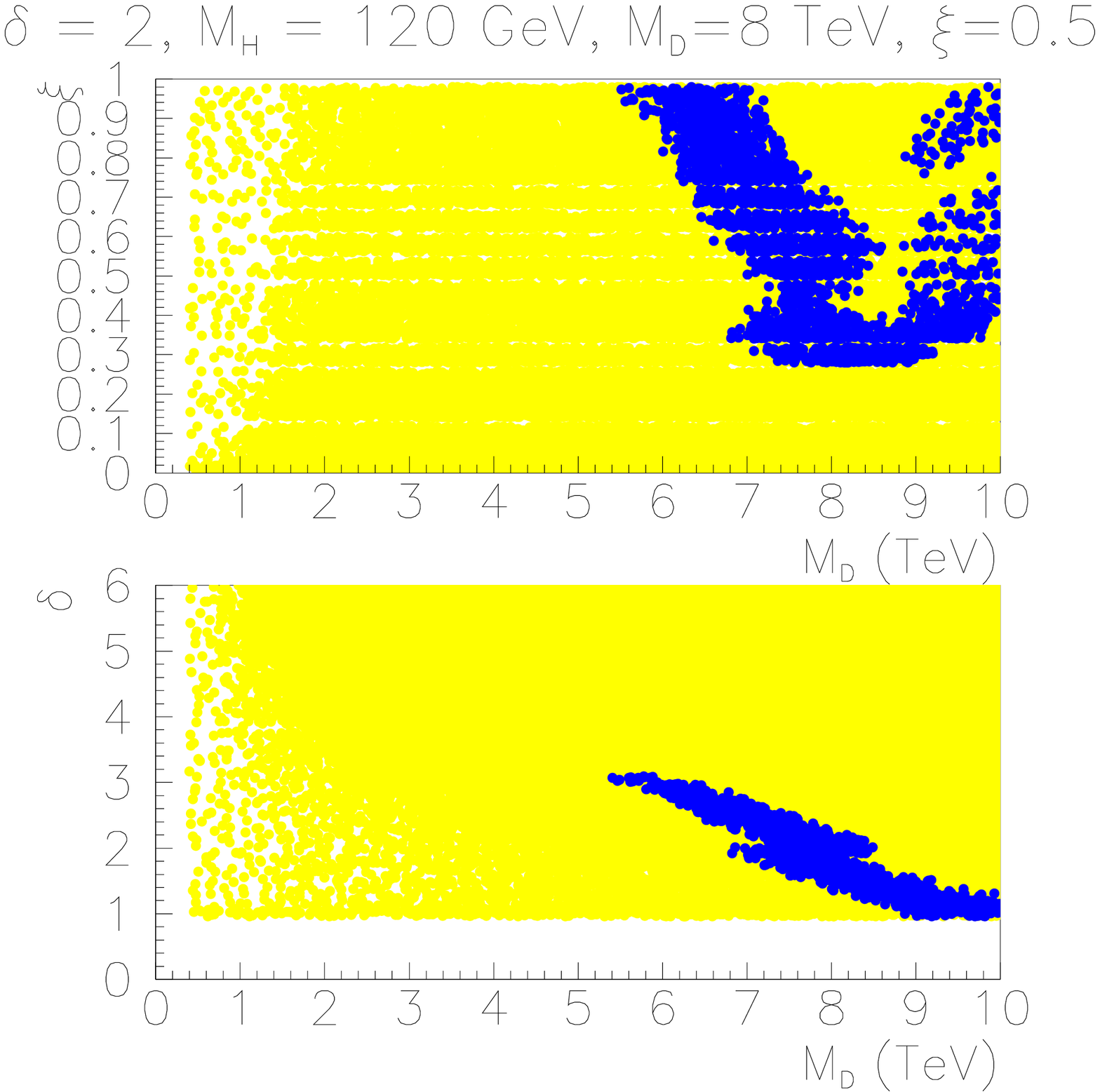}&
\includegraphics[width=4.5cm,height=6.5cm]{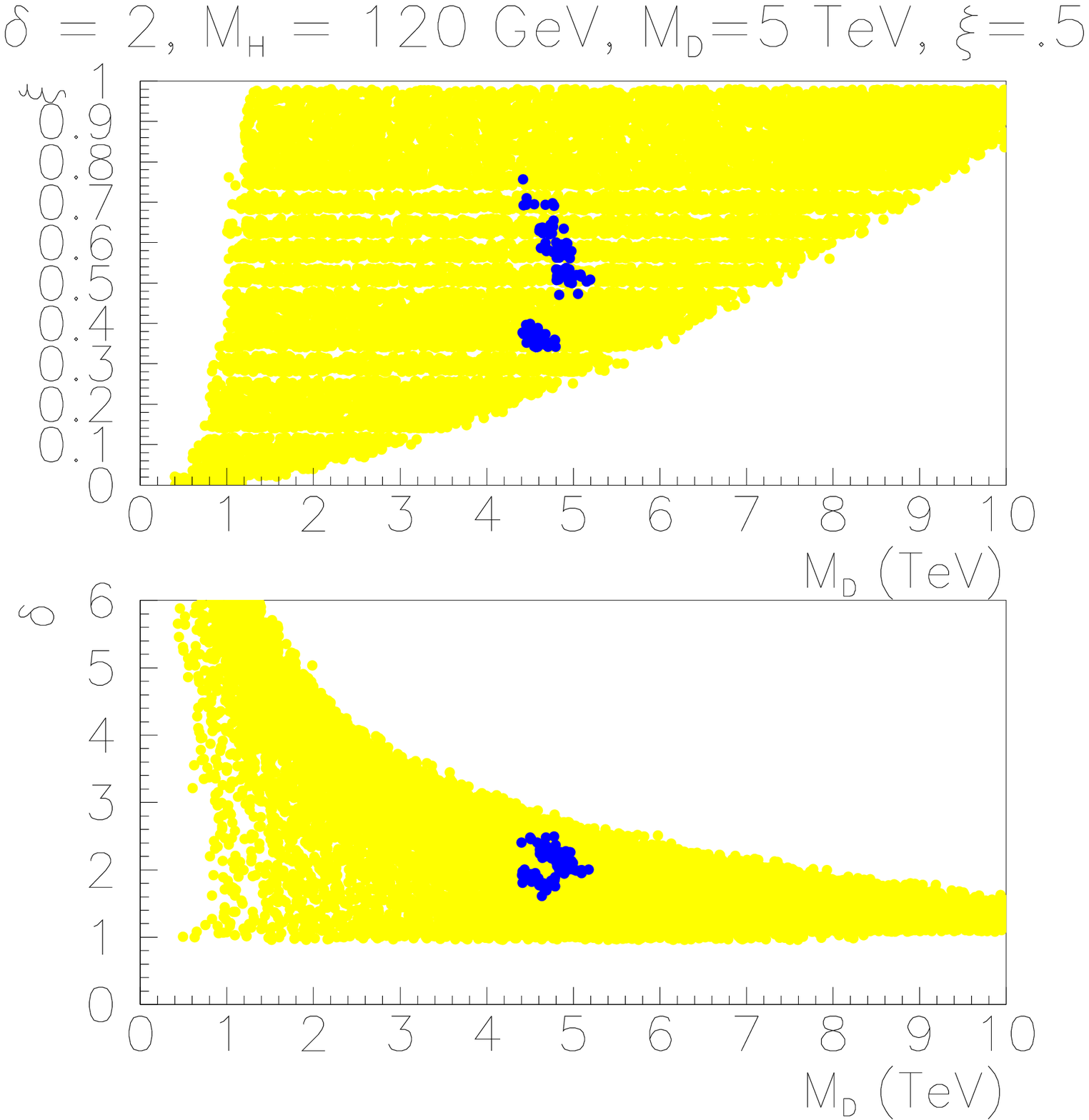}
\end{tabular}
\caption{95\% CL contours for determination of the ADD parameters,
  $\md$, $\xi$ and $\delta$ assuming $m_{\heff}=120\gev$. The plots
  are all for $\delta=2$ and $\xi=0.5$. The upper two plots and lower
  left plot are obtaining
assuming $L=100\fbi$ at the LHC, $\rts=350\gev$ Higgs measurements
at the LC, and $\rts=500\gev$ and $\rts=1000\gev$ $\gam+\etmiss$
measurements at the LC with $L=1000\fbi$ and $L=2000\fbi$ at the
two respective energies.  They are for different $\md^0$ values:
upper left ---  $\md^0=2\tev$;
upper right --- $\md^0=5\tev$;
lower left --- $\md^0=8\tev$.
The lower right plot is a repeat of the $\md^0=5\tev$ case, but
assuming lower integrated luminosities: $L=30\fbi$ at the LHC
and $L=500\fbi$ and $L=1000\fbi$ at $\rts=500\gev$ and $\rts=1000\gev$
at the LC.
The larger light grey (yellow) regions are the 95\% CL regions in
the $\xi,\md$ and $\delta,\md$ planes using only $\Delta\chi^2(LHC)$.
The smaller dark grey (blue) regions or points are the 95\% CL regions in
the $\xi,\md$ and $\delta,\md$ planes using $\Delta\chi^2(LHC+LC)$.
  }
\label{95cla}
\end{center}
%}
\end{figure}

The parameters of the model can be determined by combining several measurements
that can be performed at LHC and a LC: here we closely follow
the discussion of \cite{lavoro}.

 For the LHC Higgs signal in visible channels, we compute the $\Delta\chi^2$
for a model relative to expectations for an input model as follows.

For some central choice of parameters, define $S_0=f_0 B_0$ and ${\nsd}_0= S_0/\sqrt B_0$.  Then, $\Delta
S_0^2=S_0+B_0=B(1+f_0)=[S_0/{\nsd}_0]^2(1+f_0)$. As a result, we can compute 
$\Delta\chi^2$ for some other choice of parameters that yields signal
rate $S$ as
\beq
\Delta\chi^2={(S-S_0)^2\over \Delta S_0^2}={{\nsd}_0^2\over
    1+f_0}\left[{S\over S_0}-1 \right]^2={{\nsd}_0^2\over
    1+f_0}\left[\left(1-BR_{\heff\to invisible}\over 1-BR^0_{\heff\to invisible}\right)-1\right]^2\, .
\eeq
We obtain ${\nsd}_0$ as previously described.
In principle, $f_0$ should be computed on a channel by channel basis.
In \cite{lavoro} we have  adopted an
average value of $S_{SM}/B=f_{SM}$ for the SM Higgs rates (assuming
no invisible decays) that applies to all channels and compute
$f_0= \left(1-BR^0_{\heff\to invisible}\right)f_{SM}$. 
We have chosen $f_{SM}=0.5$, a value that we consider somewhat
conservative except for the $\gam\gam$ final state mode.  

 For the LHC Higgs signal in the invisible final state, we
  employed the detailed results of \cite{Eboli:2000ze} (used in
\cite{cmsnote}), in which the Higgs signal  and background event rates are given for
the $WW\to Higgs \to invisible$ channel assuming SM production
rate and 100\% invisible branching ratio. The background cross section
extracted from \cite{Eboli:2000ze} is $\sigma_{B_{inv}}=409.6\fb$.
Signal cross sections, $\sigma_{S_{inv}^{SM}}$, 
for 100\% invisible branching ratio are
given for Higgs masses ranging from $110\gev$ to $400\gev$.  
These cross sections are multiplied by the assumed integrated
luminosity to obtain the signal and background rates, $S_{inv}^{SM}$
and $B_{inv}$. We
rescale the signal rate  using $S_{inv}^{\heff}=BR_{\heff\to
  invisible}S_{inv}^{SM}$ and compute 
the error in the signal rate as $[\Delta S_{inv}^{\heff } ]^2=S_{inv}^{\heff}+B_{inv}$.

As we shall see, a TeV-class $e^+e^-$ linear collider will be able to
improve the determination of the ADD model parameters very
considerably with respect to the LHC alone, making use
of the Higgs signals in both visible and invisible final states
and also of the $\gam+\etmiss$ signal. 
For the $\gam+\etmiss$ signal, we have
employed the TESLA study results of \cite{grahamwilson} for the
signal. The signal cross section in Fig.~1 of \cite{grahamwilson} 
was computed  assuming 80\% $e^-$ beam polarization
and 60\% $e^+$ beam polarization, as well as cuts on the final
state photon of $E_{\gamma} < 0.625 E_{beam}$,
$|cos \theta_{\gamma} | < 0.90$ and $E_T >
0.06 E_{beam}$.
The  $e^+e^-\to \nu_e\anti \nu_e+\gam$ background has been
 computed using
the {\tt KK}  \cite{Ward:2002qq} and {\tt nunugpv}~\cite{Montagna:ec}
simulation programs. Results from the two programs agree well.
For the polarization choices and cuts listed above, we find
$\sigma_B= 102 (106.7),$125.7 (123.7), and $202.3 (195.6)\fb$
using the {\tt KK} ({\tt nunugpv}) programs 
at $\rts=1000,~800,~500\gev$,
respectively. (The $\rts=800\gev$ result is in rough agreement
with that employed in \cite{grahamwilson}.)

Figure~\ref{95cla} considers fixed input parameters of $\delta=2$ and
$\xi=0.5$; the input $M_D$ is varied between the first three
subfigures while the luminosities assumed are reduced for the fourth figure.
We observe  that the ability of the LHC to
determine the input parameters is very limited; however  
by including the precision LC data, quite precise $\delta$ and $\md$
determination is possible so long as $\md$ is not too big.  In contrast,
the precision of the $\xi$ determination leaves much to be desired in
all but the first ($\md=2\tev$) case where the invisible
branching ratio is large and the SM visible modes are suppressed
and varying rapidly as a function of $\xi$.  
Comparing the lower right figure to the upper right
figure, we see that the decline in precision resulting from lowering
the LHC and LC luminosities is not that drastic.

%\section{Conclusions}
%The ADD model is a new approach to the hierarchy problem based
%on extra dimensions. The parameters of the model are the number of extra dimensions $\delta$, the $D$-dimensional Planck mass $M_D$ and the parameter $\xi$
%which determines the mixing of the Higgs and the Ricci curvature. As a result of this mixing
%a light  Higgs acquires a sizeable invisible decay which could become
%the discovery channel of the Higgs at LHC.

\bigskip

{\small I would like to thank M. Battaglia and  J. Gunion for their
collaboration on the topics discussed in this talk. }

\bigskip
%\bibliography{inv_vienna}

\begin{thebibliography}{10}

\bibitem{Belanger:2001am}
G.~Belanger, F.~Boudjema, A.~Cottrant, R.~M. Godbole, and A.~Semenov.
\newblock The MSSM invisible Higgs in the light of dark matter and g- 2.
\newblock {\em Phys. Lett.}, B519:93--102, 2001.

\bibitem{Martin:1999qf}
Stephen~P. Martin and James~D. Wells.
\newblock Motivation and detectability of an invisibly-decaying Higgs boson at
  the Fermilab Tevatron.
\newblock {\em Phys. Rev.}, D60:035006, 1999.

\bibitem{Joshipura:1993hp}
Anjan~S. Joshipura and J.~W.~F. Valle.
\newblock Invisible Higgs decays and neutrino physics.
\newblock {\em Nucl. Phys.}, B397:105--122, 1993.

\bibitem{Belotsky:2002ym}
K.~Belotsky, D.~Fargion, M.~Khlopov, R.~Konoplich, and K.~Shibaev.
\newblock Invisible Higgs boson decay into massive neutrinos of 4th generation.
\newblock {\em Phys. Rev.}, D68:054027, 2003.

\bibitem{Arkani-Hamed:1998rs}
N.~Arkani-Hamed, S.~Dimopoulos, and G.~R. Dvali.
\newblock {\em Phys. Lett.}, B429:263--272, 1998.

\bibitem{Antoniadis:1998ig}
I.~Antoniadis, N.~Arkani-Hamed, S.~Dimopoulos, and G.~R. Dvali.
\newblock {\em Phys. Lett.}, B436:257--263, 1998.

\bibitem{Giudice:2000av}
G.~F. Giudice, R.~Rattazzi, and J.~D. Wells.
\newblock {\em Nucl. Phys.}, B595:250--276, 2001.

\bibitem{Wells:2002gq}
J.~D. Wells.
\newblock hep-ph/0205328.

\bibitem{Allanach:2004ub}
M.~Battaglia et al in B.~C. Allanach et~al.,
\newblock Les Houches 'Physics at TeV colliders 2003' Beyond the Standard Model
  working group: Summary report.
\newblock 2004.

\bibitem{lavoro}
M.~Battaglia, D.~Dominici, and J.~F. Gunion.
\newblock In preparation.

\bibitem{Giudice:2003tu}
Gian~Francesco Giudice and Alessandro Strumia.
\newblock Constraints on extra-dimensional theories from virtual- graviton
  exchange.
\newblock {\em Nucl. Phys.}, B663:377--393, 2003.


\bibitem{cmsnote}
S.~Abdullin {\it et al.}
\newblock CMS NOTE-2003/033.

\bibitem{Vacavant:2001sd}
L.~Vacavant and I.~Hinchliffe.
\newblock {\em J. Phys.}, G27:1839--1850, 2001.

\bibitem{Giudice:1998ck}
G.~F. Giudice, R.~Rattazzi, and J.~D. Wells.
\newblock {\em Nucl. Phys.}, B544:3--38, 1999.

\bibitem{schumacher}
M~Schumacher.
\newblock Note.
\newblock LC-PHSM 2003-096.




\bibitem{Eboli:2000ze}
Oscar J.~P. Eboli and D.~Zeppenfeld.
\newblock Observing an invisible Higgs boson.
\newblock {\em Phys. Lett.}, B495:147--154, 2000.


\bibitem{grahamwilson}
Graham~W. Wilson.
\newblock LC-PHSM-2001-010, Feb. 2001.

\bibitem{Ward:2002qq}
B.~F.~L. Ward, S.~Jadach, and Z.~Was.
\newblock Precision calculation for $e^+ e^- \to 2f$: The k k mc project.
\newblock {\em Nucl. Phys. Proc. Suppl.}, 116:73--77, 2003.

\bibitem{Montagna:ec}
G.~Montagna, O.~Nicrosini, and F.~Piccinini.
\newblock Nunugpv: A Monte Carlo event generator for $e^+ e^- \to \nu \bar \nu
  \gam$ events at lep.
\newblock {\em Comput. Phys. Commun.}, 98:206, 1996.



\end{thebibliography}

\end{document}